\documentclass[aps,showpacs,preprintnumbers,amsmath, amssymb]{revtex4}

\usepackage{float}
\usepackage{graphics,epsfig}
\usepackage{graphicx}
\usepackage{dcolumn}
\usepackage{bm}

\begin{document}
\baselineskip=0.8 cm

\title{{\bf Thermodynamics of general flat space and boson star quasi-local systems}}
\author{Yan Peng$^{1}$\footnote{yanpengphy@163.com},Jianjun Fang$^{2,3}$\footnote{jian-junfang@163.com}, Shuangxi Yi$^{2}$\footnote{yisx2015@qfnu.edu.cn}, Guohua Liu$^{1}$\footnote{liuguohua1234@163.com}}
\affiliation{\\$^{1}$ School of Mathematical Sciences, Qufu Normal University, Qufu, Shandong 273165, China}
\affiliation{\\$^{2}$ School of Physics and Physical Engineering, Qufu Normal University, Qufu,Shandong 273165, China}
\affiliation{\\$^{3}$ CFisUC, Department of Physics, University of Coimbra, 3004-516 Coimbra, Portugal}

\vspace*{0.2cm}
\begin{abstract}
\baselineskip=0.6 cm
\begin{center}
{\bf Abstract}
\end{center}

We study thermodynamics of flat space/boson star systems enclosed in a
scalar reflecting box with St$\ddot{u}$ckelberg mechanism.
We also disclose effects of model parameters on
transitions and the properties appear to be qualitatively the
same as those in holographic St$\ddot{u}$ckelberg transitions.
Moreover, we obtain a relation $\bar{\zeta}\thickapprox 2 \tilde{\zeta}$
and the second order characteristic exponent, which also hold in
holographic superconductor theories.
The similarity between quasi-local thermodynamic transitions and holographic transitions
provides additional evidences that
holographic theories may also exist in the quasi-local space.

\end{abstract}

\maketitle
\newpage
\vspace*{0.2cm}

\section{Introduction}

According to the AdS/CFT correspondence,
strongly interacting physics on the boundary
can be holographically described by weakly coupled bulk AdS
gravity theories \cite{Maldacena,Gubser,Witten}.
And the bulk thermodynamic transitions were usually applied to study
the AdS/CFT correspondence \cite{S.A. Hartnoll}-\cite{J.P. Gauntlett}.
However, it was argued that the holography first discovered in the AdS spacetimes
may also exist in other confined spaces based on the fact that
thermodynamic transitions in a reflecting box are strikingly similar to those in AdS gravities
\cite{S. Carlip,Andrew,J. X. Lu}.

Along this line, it is interesting to further research on the similarity
between quasi-local thermodynamic transitions in a box and transitions in the AdS gravity.
For a single Maxwell field coupled to the gravity, it was found that quasi-local
thermodynamic transition structures of bulk gravity systems are similar to cases in
the asymptotically AdS background \cite{Robert,P. Hut,Gibbons}.
Lately, this box gravity system was further generalized
by including an additional scalar field coupled to the Maxwell field \cite{Pallab Basu}.
In particular, the phase diagram of this complete box gravity system
with a scalar reflecting boundary is strikingly similar to
that of holographic transitions.
We further generalized the box confined transition model by considering nonzero scalar mass and
St\"{u}ckelberg mechanism with higher scalar field terms $\psi^{2}+\zeta\psi^{4}$
and $\zeta$ as the model parameter \cite{Yanpeng-1,Yanpeng-2}.
We showed the similarity between the box gravity and the AdS gravity through effects of model parameters on
the threshold chemical potential and also the order of phase transitions.
Recently, a new St\"{u}ckelberg mechanism with higher correction terms of the
form $\psi^{2}+\zeta\psi^{6}$ was discussed in the holographic superconductor theory
and this new type of St\"{u}ckelberg mechanism can trigger richer physics compared with results of usual
holographic St\"{u}ckelberg superconductor models \cite{R.-G. Cai,Yanpeng-4}.
In particular, it admits a new type of first order transition between scalar hairy states.
So it is meaningful to further generalize the box gravity by considering this new St$\ddot{u}$ckelberg mechanism
and disclose similarities between the box confined system and the AdS gravity.

We put the rest part of this work as follows. In section II, we introduce a
flat space and boson star system in a reflecting box with St$\ddot{u}$ckelberg mechanism.
In section III, we study thermodynamics of the box gravity system
and also search for the similarity between quasi-local transitions and holographic transitions.
The conclusion will be presented at the end of this work.

\section{Equations of motion and boundary conditions}

In this work, we are interested in the flat space and boson star transition model
enclosed in a scalar reflecting box with radius labeled as $r=r_{b}$.
And the general St\"{u}ckelberg Lagrange density with a scalar field and a Maxwell field
coupled in the asymptotically flat spacetime is \cite{R.-G. Cai,Yanpeng-4}:
\begin{eqnarray}\label{lagrange-1}
\mathcal{L}=R-[F^{\alpha \beta}F_{\alpha \beta}-\partial^{\alpha} \psi \partial_{\alpha} \psi-G(\psi)(\partial \theta
-q A_{\alpha})^{2}-m^{2}\psi^{2}].
\end{eqnarray}
Here, the scalar field $\psi$ is only radial dependence as
$\psi=\psi(r)$ with mass $m$ and charge q.
We also take the Maxwell field with only nonzero t component in the form
$A=\phi(r)dt$. We consider the transition model with the general function of the form
$G(\psi)=\psi^{2}+q^4 \zeta\psi^{6}$, where $\zeta$ is the St\"{u}ckelberg
parameter and $\zeta=0$ is the case without St$\ddot{u}$ckelberg mechanism \cite{Pallab Basu,R.-G. Cai,Yanpeng-4}.
We mention that this new type of St\"{u}ckelberg mechanism
usually leads to richer physics compared with transition
models in \cite{Yanpeng-1} with a function $G(\psi)=\psi^{2}+q^2 \zeta\psi^{4}$
and $\zeta$ as the model parameter.
In the calculation, we can take $\theta=0$ according to the symmetry
$A_{\mu}\rightarrow A_{\mu}+\partial \alpha$ and $\theta\rightarrow \theta+\alpha$.

The spherical symmetric boson star metric in the asymptotically flat gravity can be putted in the form
\begin{eqnarray}\label{AdSBH}
ds^{2}&=&-g(r)h(r)dt^{2}+\frac{dr^{2}}{g(r)}+r^{2}(d\theta^{2}+sin^{2}\theta d\varphi^{2}).
\end{eqnarray}
We impose $g(\infty)=1$ and $h(\infty)=1$ to recover the flat space at the infinity.

Equations of metrics and matter fields are
\begin{eqnarray}\label{BHpsi}
\frac{1}{2}\psi'(r)^{2}+\frac{g'(r)}{rg(r)}+\frac{q^2G(\psi)\phi(r)^2}{2g(r)^2h(r)}+\frac{\phi'(r)^2}{g(r)h(r)}-\frac{1}{r^2g(r)}+\frac{1}{r^2}+\frac{m^2}{2g}\psi^2=0,
\end{eqnarray}
\begin{eqnarray}\label{BHpsi}
h'(r)-rh(r)\psi'(r)^2-\frac{q^2rG(\psi)\phi(r)^2}{g(r)^2}=0,
\end{eqnarray}
\begin{eqnarray}\label{BHphi}
\phi''+\frac{2\phi'(r)}{r}-\frac{h'(r)\phi'(r)}{2h(r)}-\frac{q^2G(\psi)\phi(r)}{2g(r)}=0,
\end{eqnarray}
\begin{eqnarray}\label{BHg}
\psi''(r)+\frac{g'(r)\psi'(r)}{g(r)}+\frac{h'(r)\psi'(r)}{2h(r)}+\frac{2\psi'(r)}{r}+\frac{q^2G'(\psi)\phi(r)^2}{2g(r)^2h(r)}-\frac{m^2}{g}\psi=0,
\end{eqnarray}
where we define $G'(\psi)=\frac{d G(\psi)}{d \psi}=2 \psi+ 6 q^4 \zeta \psi^{5}$.

With numerical methods, we integrate the equations from $r=0$ to $r=r_{b}$
to obtain solutions with the scalar field vanishing at the box boundary.
We put asymptotical behaviors of fields around $r=0$ in the form
\cite{Pallab Basu}
\begin{eqnarray}\label{InfBH}
&&\psi(r)=a_{1}+b_{1} r^2+\cdots,\nonumber\\
&&\phi(r)=a_{2}+b_{2} r^2+\cdots,\nonumber\\
&&g(r)=1+a_{3} r^2+\cdots,\nonumber\\
&&h(r)=a_{4}+b_{4} r^2+\cdots,
\end{eqnarray}
Putting relations (7) into equations (3-6) and considering leading
terms, we find that the solutions can be determined by parameters
$a_{1}$, $a_{2}$ and $a_{4}$.
We can also take the boundary as $r_{b}=1$ using the rescaling $r\rightarrow \kappa r$.
And the scalar fields close to the box behaves as
\begin{eqnarray}\label{InfBH}
\psi\rightarrow \psi_{1}+\psi_{2}(1-r)+\cdots.
\end{eqnarray}
Since the scalar field vanishes at the box boundary,
we impose $\psi_{1}=0$ and show that transitions can be described
by the other parameter $\psi_{2}$ similar to methods in holographic
superconductor theories \cite{Yanpeng-1}.
And the value of the vector field $\phi(r)$ on the box $\mu=\phi(1)$ is the chemical potential.
We can also transform the solutions to satisfy $g_{tt}(1)=-1$
according to the symmetry $h\rightarrow \gamma^2h,~\phi\rightarrow \phi,~t\rightarrow\frac{t}{\gamma}$ \cite{Pallab Basu}.

\section{Thermodynamics of the quasi-local gravity system}

In this part, we firstly disclose properties of phase transitions from behaviors of the free energy.
We plot the free energy with respect to the chemical potential in Fig. 1 with
$q=100$, $m^{2}=-2$ and different values of $\zeta$.
For the case of $\zeta=0,~0.004 ~or~ 0.006$ in $(a),~(b)$ and $(c)$ of Fig. 1,
we find that the free energy is smooth around the critical chemical potential,
which corresponds to the second order flat space/boson star transition.
Besides the second order transition, $(c)$ shows that there is also a
new first order phase transition between the boson star phases
at $\tilde{\mu}_{c}= 0.02925$.
When we choose a larger parameter $\zeta=0.01$ in $(d)$ of Fig .1,
there is only the first order flat space/boson star phase transition
at the threshold value $\mu_c=0.02793$.
We mention that phase structures of this quasi-local gravity system with various $\zeta$
are qualitatively the same as those of holographic St$\ddot{u}$ckelberg transitions
where large enough St$\ddot{u}$ckelberg parameter could trigger an additional first order transition
between hairy states and for very large St$\ddot{u}$ckelberg parameter,
there is only the first order hairless state/hairy state transition.

\begin{figure}[h]
\includegraphics[width=185pt]{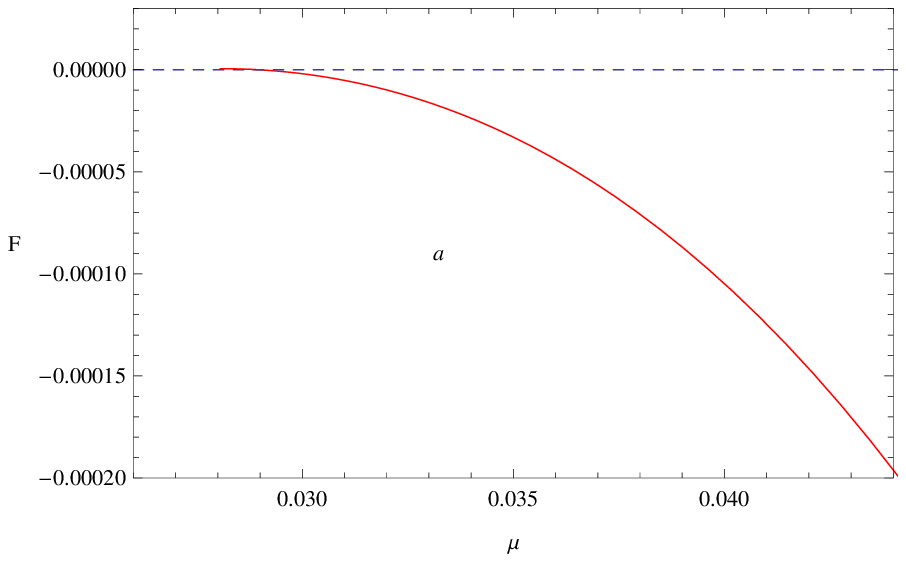}\
\includegraphics[width=190pt]{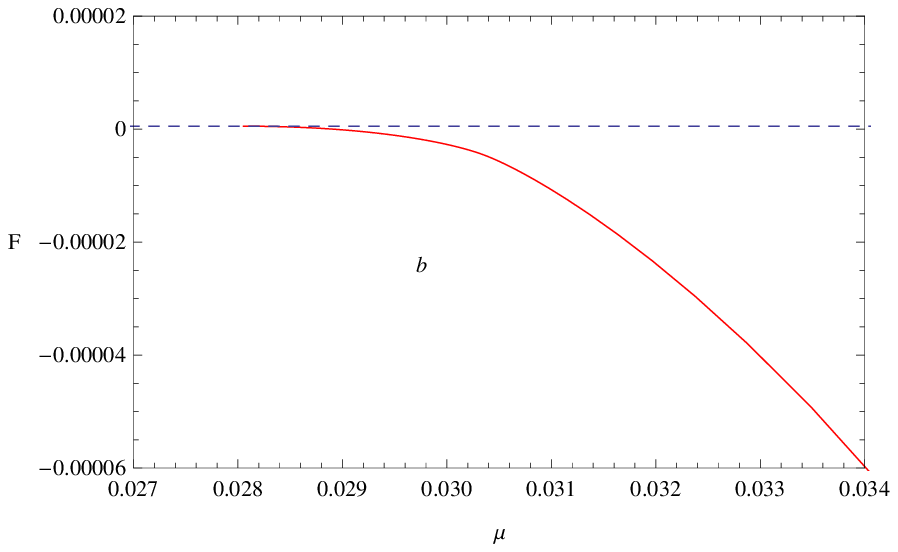}\
\includegraphics[width=190pt]{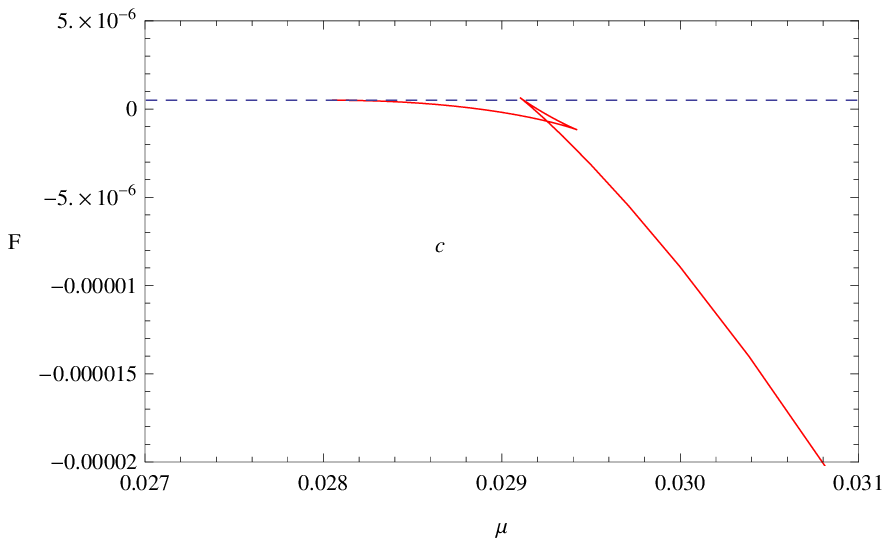}\
\includegraphics[width=190pt]{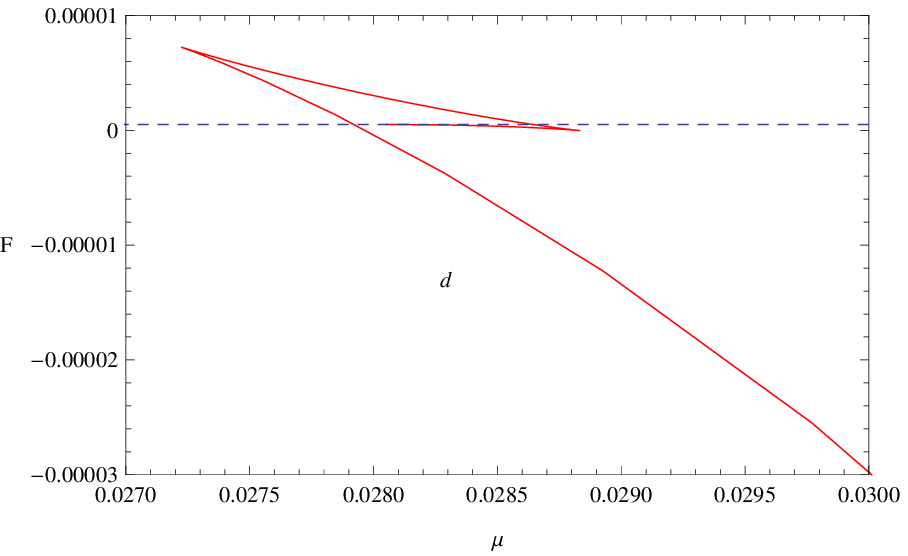}\
\caption{\label{EEntropySoliton} (Color online) The free energy with respect to
the chemical potential $\mu$ in cases of $q=100$, $m^{2}=-2$ and various $\zeta$: $(a)$ $\zeta=0$,~
$(b)$ $\zeta=0.004$,~$(c)$ $\zeta=0.006$ and $(d)$ $\zeta=0.01$.
The dashed lines show the free energy of
the flat space and red solid curves are with the boson star phases.}
\end{figure}

We should point out that the free energy here corresponds to the energy in the gravitational box
with a background subtraction of the pure Minkowski box \cite{Pallab Basu}. So curves in Fig. 1 reveal
the property of bulk transitions. We further applied approaches similar to
holography and found that operators on the hard cut-off box boundary covers some
information of the bulk gravity \cite{Yanpeng-1,Yanpeng-2}.
In this more general model with richer phase diagrams, we will show that operators
on the boundary can still be used to disclose the critical transition points
and the order of bulk transitions.

Now we study bulk transitions from behaviors of the box boundary
operator $\psi_{2}$ similar to the usual holographic approaches
in the AdS gravity. We plot $\psi_{2}$ as a function of the chemical potential in Fig. 2.
From the panels $(a)$, $(b)$ and $(c)$ of Fig. 1 and Fig. 2, we see that $\psi_{2}$ is continuous
around the second order transition point $\mu=0.02805$.
There is also a jump of the operator $\psi_{2}$ at $\mu= 0.02925$
in $(c)$ of Fig. 2 corresponding to the first order transition in $(c)$ of Fig.1.
And in $(d)$ of Fig.2 with $\zeta=0.01$,
$\psi_{2}$ has a jump at $\mu=0.02793$
corresponding to the first order transition in (d) of Fig. 1.
So the box boundary operator $\psi_{2}$
can be used to disclose the critical phase transition points
and also the order of transitions.
We conclude that the operator on the box boundary covers
part information of the bulk transition.

\begin{figure}[h]
\includegraphics[width=180pt]{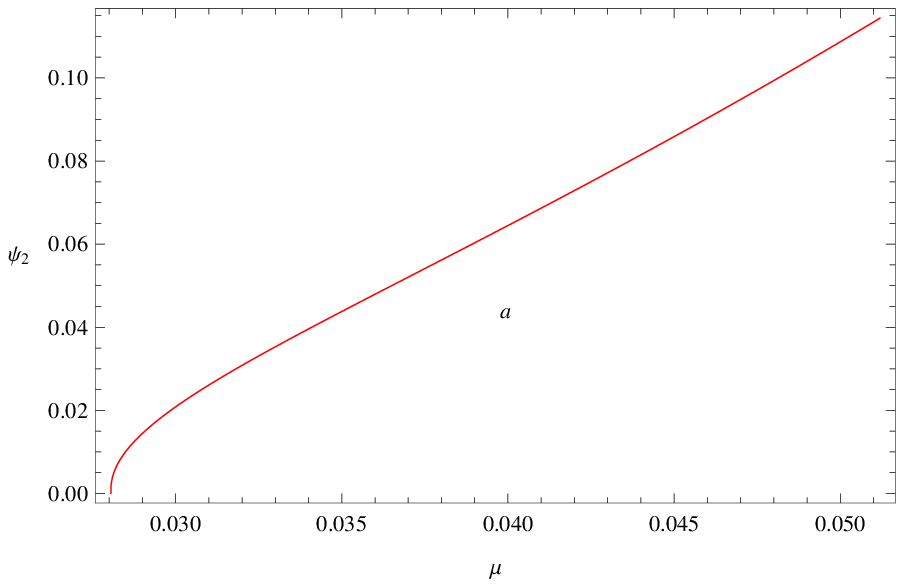}\
\includegraphics[width=180pt]{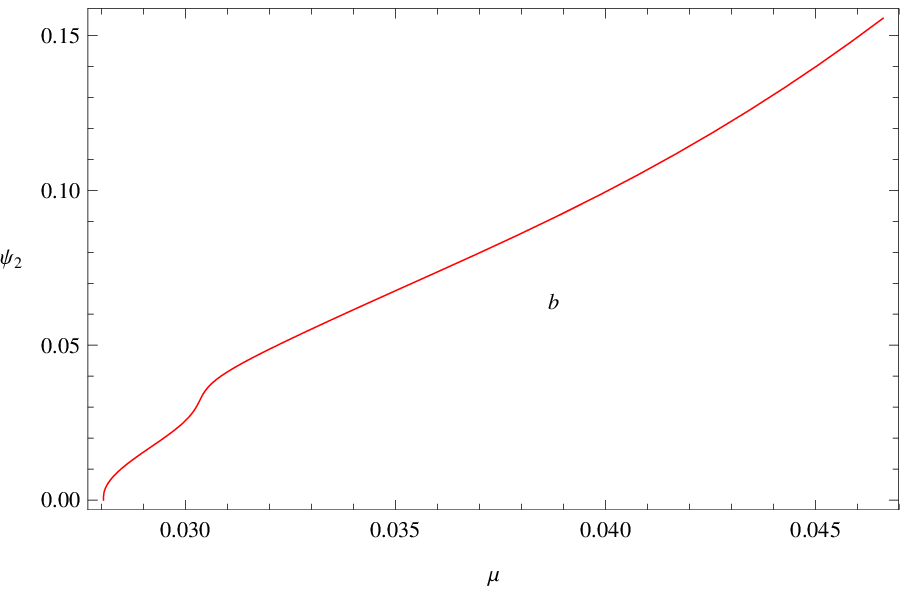}\
\includegraphics[width=180pt]{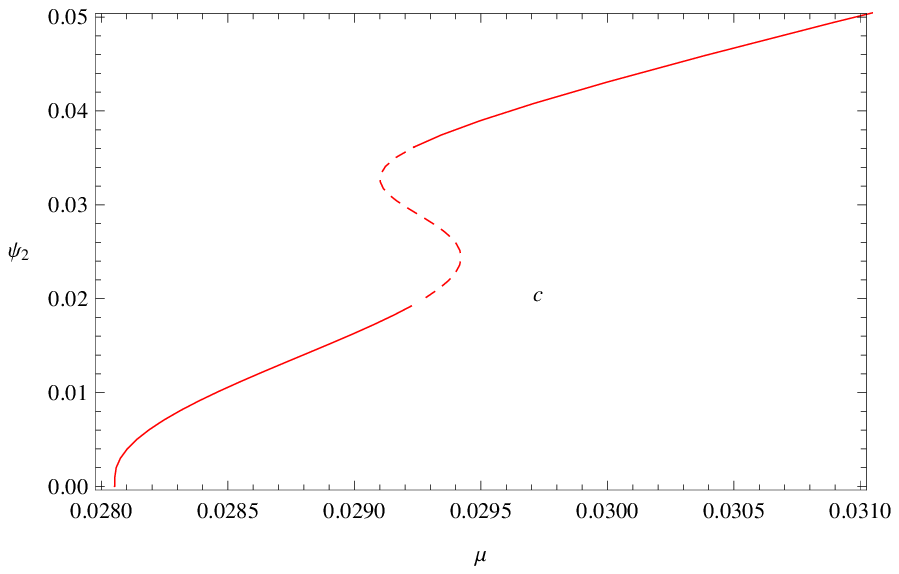}\
\includegraphics[width=180pt]{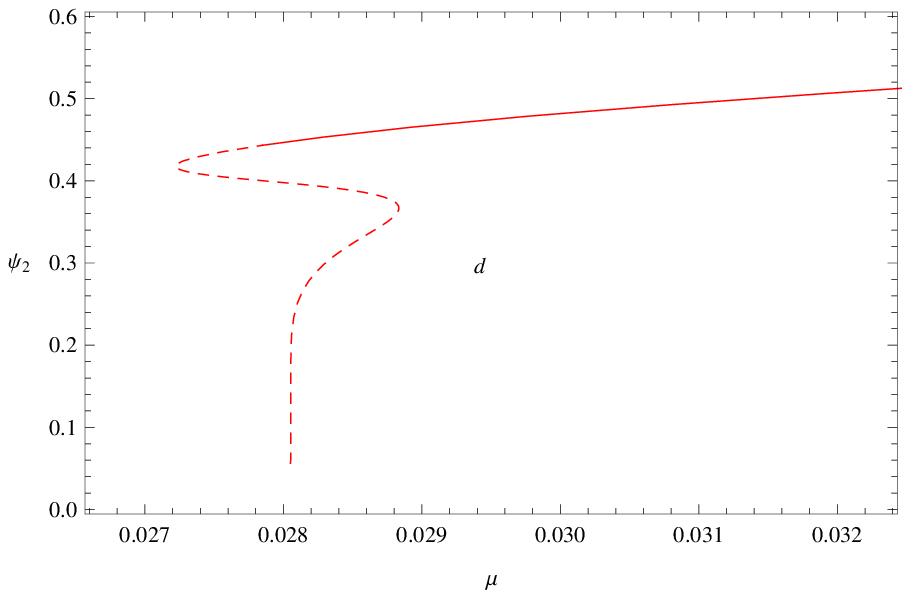}\
\caption{\label{EEntropySoliton} (Color online) The box boundary operator $\psi_{2}$ with respect
to the chemical potential $\mu$ in cases of $q=100$, $m^{2}=-2$ and  various $\zeta$ as:
$(a)$ $\zeta=0$,~$(b)$ $\zeta=0.004$,~$(c)$ $\zeta=0.006$ and $(d)$ $\zeta=0.01$.
The red solid lines represent the thermodynamically stable boson star phase and
the red dashed curves correspond to the unstable boson star phase. }
\end{figure}

In the AdS gravity, there is a characteristic
exponent $\frac{1}{2}$ corresponding to the second order
holographic transition \cite{Yanpeng-4,Rong-Gen Cai}.
And this second order characteristic
exponent $\frac{1}{2}$ also exists in box gravity systems
\cite{Yanpeng-1,Yanpeng-2}.
Along this line, it is interesting to examine
whether this property also hold in our general
box gravity model with St$\ddot{u}$ckelberg mechanism.
In fact, we obtain the relation $\psi_{2}\thickapprox0.430(\mu-\mu_{c})^{1/2}$
with $\mu_{c}=0.02805$ by fitting the numerical data.
It can be easily seen from Fig. 3 that the blue solid line
representing the fitted formula almost coincides
with solid red lines obtained from numerical data
around the phase transition points.
Here we conclude that the second order characteristic
exponent also exist in the box gravity with St$\ddot{u}$ckelberg
mechanism.

\begin{figure}[h]
\includegraphics[width=200pt]{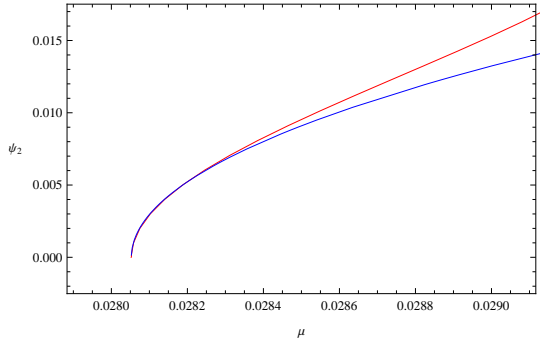}\
\caption{\label{EEntropySoliton} (Color online) We show behaviors
of $\psi_{2}$ as a function of the chemical potential with $q=100$, $m^{2}=-2$ and  $\zeta=0.006$.
The solid red line represents boson star phases and the solid blue line corresponds to
$\psi_{2}\thickapprox0.430(\mu-\mu_{c})^{1/2}$ with $\mu_{c}=0.02805$.}
\end{figure}

With more detailed calculations, we can define
threshold values $\bar{\zeta}$ and $\widetilde{\zeta}$
to describe phase structures.
When $\zeta\leqslant\tilde{\zeta}$, there is only the second order
flat space/boson star transition at the critical chemical potential $\mu_{c}$.
In the case of $\tilde{\zeta} <\zeta<\bar{\zeta}$,
there are new types of first order transitions
between boson star phases at $\tilde{\mu}_{c}$
besides the second order flat space/boson star transition at $\mu_c$.
And for $\zeta\geqslant \bar{\zeta}$,
only a first order flat space/boson star transition appears.

In holographic superconductor transitions,
a relation $\bar{\zeta}\thickapprox 2 \tilde{\zeta}$
holds \cite{Yanpeng-4} and more negative scalar mass
corresponds to a smaller critical chemical potential of second order transitions \cite{Q. Pan,YQ}.
Now we extend the discussion to box gravity system.
In the left panel of Fig. 4,
we show values of $\tilde{\zeta}$ and $\bar{\zeta}$ as a function of $m^{2}$ with $q=100$ and
arrive at the same relation $\bar{\zeta}\thickapprox 2 \tilde{\zeta}$.
And in the right panel of Fig. 4,
we show that more negative scalar mass $m^{2}$
corresponds to a smaller critical chemical potential $\mu_{c}$.
These properties provide more similarity between the
confined transition and holographic transitions.

\begin{figure}[h]
\includegraphics[width=200pt]{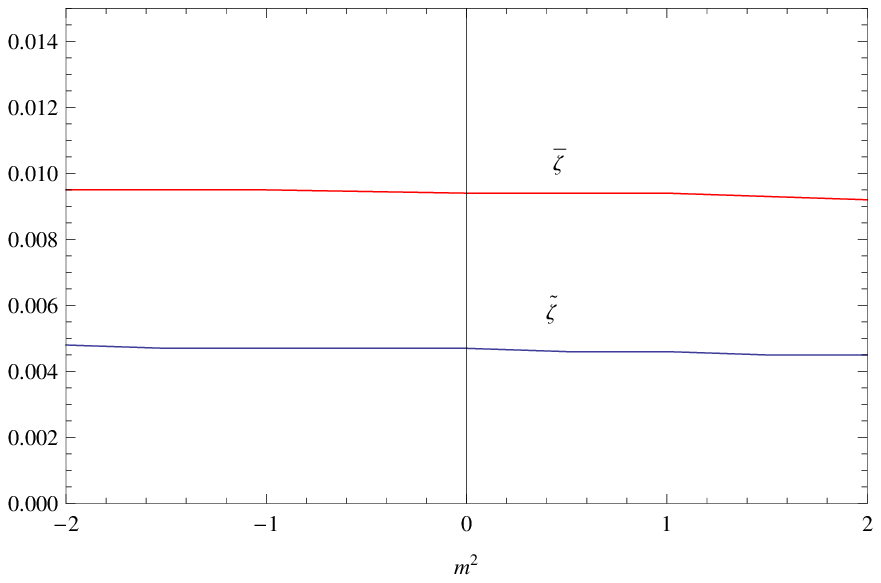}\
\includegraphics[width=210pt]{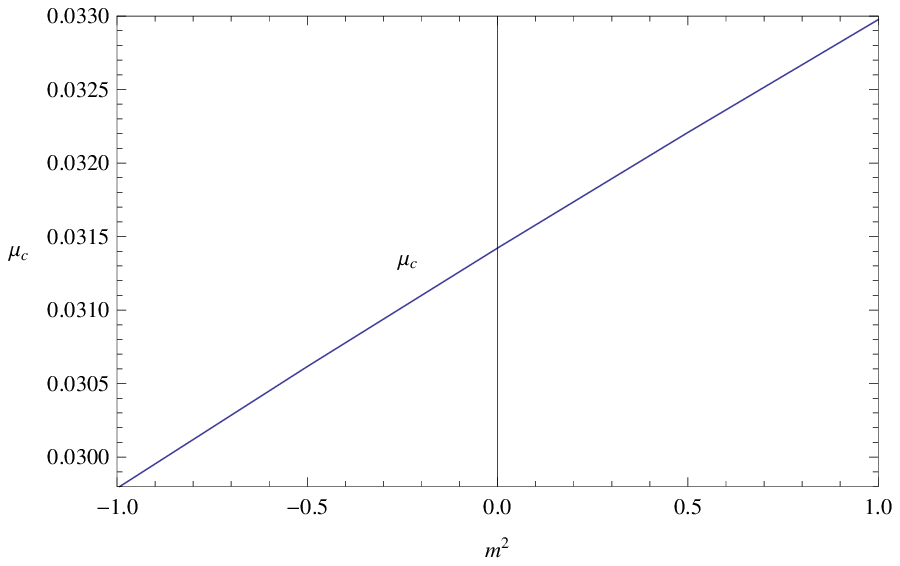}\
\caption{\label{EEntropySoliton} (Color online) The left panel shows effects of the scalar mass on the
critical parameter $\overline{\zeta}$ (red) and  $\widetilde{\zeta}$ (blue) with $q=100$.
And in the right panel, the blue line is the critical chemical potential $\mu_{c}$
as a function of the scalar mass $m^2$ with $q=100$ and $\zeta=0.004$. }
\end{figure}

We study thermodynamics of the scalar field and the box gravity system.
And we mention that the dynamical properties of this quasi-local
model have been analyzed in \cite{box1,box2,box3,box4,box5,box6,box7}.
It seems that the scalar reflecting box serves as a way
to confine the scalar field and make scalar hair easier
to form outside the black hole horizon.
Another well known way to confine the scalar field
is adding an AdS boundary where a potential well in the near horizon
region forms due to the AdS boundary \cite{sn1,sn2}.
Moreover, there is no scalar hair theorem in regular neutral reflecting stars \cite{Hod1,SBS}
and static scalar fields can condense around charged reflecting stars
\cite{Hod2,Hod3,Hod4,EMP,YP1,YP2}.
So it is also very interesting to extend the discussion to the reflecting star
with box boundary conditions.

\section{Conclusions}

We investigated thermodynamics of a flat space/boson star model
with St\"{u}ckelberg mechanism in the asymptotically flat quasi-local gravity.
We found that the operator $\psi_{2}$ on the box boundary is useful in revealing the
critical phase transition points and also the order of transitions in this general model.
In particular, we obtained the relation
$\psi_{2}\varpropto(\mu-\mu_{c})^{1/2}$, which is exactly the
characteristic property of second order holographic transitions in the AdS gravity.
Moreover, we also arrived at a relation $\bar{\zeta}\thickapprox 2\tilde{\zeta}$, which is
the same as results in St\"{u}ckelberg holographic superconductor theories.
These properties mean that the box boundary operator covers part
information of the bulk transition and it provides additional evidences that
holographic theories may also exist in the quasi-local space.

\begin{acknowledgments}

This work was supported by the Shandong Provincial Natural Science Foundation of China under Grant
Nos. ZR2018QA008 and ZR2017BA006.
This work was alo supported by the National Natural Science Foundation of China under Grant No. 11703015.

\end{acknowledgments}

\end{document}